\documentclass[11pt,twoside]{article}

\usepackage{asp2006}

\begin{document}
\title{On Flare Driven Global Waves}   
\author{Karoff, C.}   
\affil{School of Physics and Astronomy, University of Birmingham, Edgbaston, Birmingham B15 2TT, UK
\\
Danish AsteroSeismology Centre (DASC), Department of Physics and Astronomy, University of Aarhus, DK-8000 Aarhus C, Denmark}    

\begin{abstract} 
We recently presented evidence of a strong correlation between the energy in the high-frequency part of the acoustic spectrum of the Sun and the solar X-ray flux \citep{karoff2008}. The discovery indicates that flares drive global oscillations in the Sun in the same way that the entire Earth is set ringing for several weeks after a major earthquake, such as the 2004 December Sumatra-Andaman one. If this indication turns out to be true we might be able to use the relation between flares and the energy in the high-frequency part of the acoustic spectrum to detect e.g. flares on the far side of the Sun and flares on other solar-like stars. But, the discovery also opens many new questions such as why is it only the high-frequency part of the acoustic spectrum that is correlated with the X-ray flux? And, are there energy enough in solar flares do drive global oscillations? 
\end{abstract}

\section{Introduction}
In 1972 it was suggested by Wolff \citep{1972ApJ...176..833W} that solar flares could stimulate free oscillations in the Sun. The general idea by Wolff was that the flares would cause a thermal expansion that would act as a mechanical impulse by causing a wave of compression to move subsonically into the solar interior. Later the detection of global low-degree p-mode oscillations was confirmed in disk-interegrated sunlight \citep[by e.g.][]{1976Natur.259...92B, 1981Natur.293..443C} and soon it was generally accepted that the global p-mode oscillations were driven by stochastic excitation from granulation \citep{1988ApJ...326..462G} and not by solar flares. 

Although the p~modes seemed to be excited by the near-surface convection zone a number of studies have investigated if some parts of the excitation of the low-degree p~modes were caused by e.g. flares. These studies have come out with contradicting results. \citet{1999MNRAS.303L..63G} found a high correlation between temporally varying p-mode power measured at low degree in GONG data and the coronal mass ejection event number, but due to the way they normalized the correlation coefficient and because of the selective selection of the events it is not possible to make a quantitative evaluation of the correlation coefficient obtained by \citet{1999MNRAS.303L..63G} nor is it possible to compare their value of the correlation coefficient with values obtained in other data sets. \citet{2006SoPh..238..219A} on the other hand found no correlation between a longer disk-integrated GONG data set and flare and coronal mass ejection indices. In this study the correlation coefficients were properly normalized and there were no obvious selection effects. Analysis of BiSON data have also revealed that though the strength of the p~modes follows the distribution expected from stochastic excitation by near-surface convection, there is evidence of a few more very large events than the numbers predicted by the theory, but these events show only a poor correlation with flares \citep{1995soho....2..335C}. \citet{1998A&A...339..261F} and \citet{1998A&A...330..341F} made a statistical analysis similar to the one by \citet{1995soho....2..335C} and found a mean correlation between p~modes of degree 0 and 1 of 0.6 \% in GOLF data obtained from 1996-97 and 10.7 $\pm$ 5.9 \% in IPHIR data obtained in 1988 which could suggest a relation between the p~modes and transient events; but no matches between the events in the p~modes and in the flares or coronal mass ejections were made. \citet{1992ApJ...394L..65B} analyzed the correlation between acoustic energy and activity, not in time, but in space and found that while the energy of the p-modes with frequencies between 2.5 and 4.0 mHz is suppressed in active regions the energy in the 5.5 to 7.5 mHz frequency range is enlarges around the active regions.   

The discovery by \citet{karoff2008} of a high correlation between the energy in the high-frequency part of the acoustic spectrum and the solar X-ray flux was therefore the first evidence that supported Wolffs idea. The reason why \citet{karoff2008} were able to see a correlation where others had failed was that \citet{karoff2008} did not look at the frequency range where most of the p-mode energy is positioned around 3 mHz. Instead they looked at frequencies higher than the atmospheric acoustic cut-off frequency (5.3 mHz). Why the energy in the acoustic spectrum is more correlated with the solar X-ray flux above the cut-off frequency than below is one of the questions that we do not have a definitive answer for. We will come back to this question, but let us start by describing what we know about the global waves in the Sun with frequencies higher than the atmospheric acoustic cut-off frequency.

\section{Why high-frequency waves?}
High-frequency waves were first discovered in high-degree observations of the Sun from the Big Bear Solar Observatory \citep{1988ApJ...334..510L} and later in GOLF low-degree disk-integrated observations \citep{1998ApJ...504L..51G} . Recently they have also been seen in the BiSON disk-integrated radial-velocity data \citep{2003ESASP.517..247C} and the VIRGO intensity data \citep{2005ApJ...623.1215J}.  

Two models have been proposed in order to account for the low-degree high-frequency waves with frequencies higher than the atmospheric acoustic cut-off frequency. The first model which was originally proposed by \citet{1991ApJ...375L..35K} explains the high-frequency waves as an interference phenomenon between ingoing and outgoing waves from a localized source just beneath the photosphere. The other model which was originally proposed by \citet{1990ApJ...362..256B} suggests that the high-frequency waves are (partly) reflected by the sudden change in the temperature at the transition region between the chromosphere and the corona, and not at the photosphere which is the case for the ordinary p~modes. The later model is supported by observations of partial reflection of waves at the transition regiont by the use of time-distance helioseismology \citep{1997ApJ...485L..49J}. In either case the amount of energy that is stored in the high-frequency waves is extremely low compared to the amount of energy stored in the p-mode oscillations with frequencies below the atmospheric acoustic cut-off frequency.
 
So with this in mind we can think of three suggestions as to why the energy in the acoustic spectrum is more correlated with the solar X-ray flux above the cut-off frequency than below it:

First, as noted by \citet{karoff2008} the observed energy per mode at high frequency is significantly lower than the energy per mode observed at the peak (3 mHz) and according to \citet{2001ApJ...546..585S} one finds the p-mode energy to decrease inversely as the fourth power of the frequency (above 3 mHz) simply as a consequence of the granulation power decreasing as $\nu^{-4}$. Therefore a single event (such as a flare) that excites a mode will have a larger relative effect at high frequency because the other excitation sources are much smaller.

Secondly, the intensity signal from VIRGO is not only sensitive to oscillations in the photosphere, but also to oscillations in the chromosphere. So what we are seeing might be oscillations in the chromosphere rather than oscillations below the photosphere. This idea is supported by the fact that it is not clear how much of the energy correlated with the solar X-rays goes in to the background at high frequency and how much goes in to the distinct modes. On the other hand chromosphere oscillations are believed to manifest themselves at high frequency as a few modes with large mode line-widths \citep{1993ASPC...42..111H, 1995ASPC...76..303D} though there are not many observations of disk-integrated chromosphere oscillations. A few chromosphere modes with large line-widths could be what is seen in the analysis by \citet{karoff2008} and it indeed agrees nicely with the results from MDI and GONG by Kumar (these proceedings p. ???). The idea of chromospheric oscillations is further supported by the fact that flare models seem to suggest that most of the flare energy is released in the upper part of the chromosphere and not close to the photosphere \citep{1972SoPh...24..414H} .

And thirdly, if it is assumed that the high-frequency waves are partly reflected by the sudden change in the temperature at the transition region between the chromosphere and the corona, as suggested by \citet{1990ApJ...362..256B}, then the top of the chromosphere would be the place where the high-frequency waves would be most sensitive to excitation while p-mode oscillations with frequencies below the atmospheric acoustic cut-off frequency would not be influenced as strongly by flare energy release at the top of the chromosphere. Also, it is easy to imagine that a flare could change the temperature profile over the transition region for a while, which would have a large influence on the reflection and thus the amplitudes of the high-frequency waves in the \citet{1990ApJ...362..256B} model.

In general the two last suggestions have the advantage that the amount of energy that needs to be supplied by the flares is much smaller than if the nature of the high-frequency waves had been the same as that of the ordinary p modes. This is to some extent also true for the first suggestion as the amplitudes of the high-frequencies waves are much smaller than those of the ordinary p modes.

We are of course investigating how to discriminate between the three possible solutions to the question above. We are investigating the possibility to use observations of high-frequency waves obtained at different wavelengths in order to analyze phase and amplitude differences of high-frequency waves at different heights in the solar atmosphere. We might then be able to use these phase and amplitude differences to discriminate different the models from one and another. The observations that could be used for this are e.g. the Magneto-Optical Filters at Two Heights (MOTH) observations from the South Pole \citep{2004SoPh..220..317F}, observations with the Global Oscillations at Low Frequency New Generation (GOLF-NG) instrument at at the Observatorio del Teide (Salabert, p. ??? in these proceedings) and observations of the blue sky by the Stellar Observations Network Group (SONG) instruments (Chrisenten-Dalsgaard, p. ??? in these proceedings). SONG has the advantage the it will make high-cadence high-resolution spectra available in a large frequency range which will make it possible to study a large number of absorption lines covering a large range of heights in the solar atmosphere.

\section{Flares on the far-side of the Sun}
As solar flares impact the Earth's environment and can be highly damaging for orbiting spacecraft \citep{2006Natur.441..402C} huge efforts are conducted in order to forecast the appearance of them -- the so-called space weather discipline. The major way to do this forecast is to predict the activity on the far side of the Sun which allows a warning of up to half the solar rotation period. This is mainly done with two methods: helioseismic holography and by monitoring sky reflected Lyman $\alpha$ radiation.

The concept of helioseismic holography was first proven by \citet{2000Sci...287.1799L}. In summery the method treats the Sun as a gigantic lens that allows us to see though the Sun. This can be accomplished as helioseismic holography do not study light, but sound waves to which the Sun is transparent. In this way helioseismic holography studies the acoustic waves in a localized region on the near side of the Sun (the pupil) in order to image the acoustic waves in localized regions on the far side (the focal point). The active regions on the solar surface (e.g. sun spots) change the nature of the acoustic waves that travel though them. Therefore active regions can be seen on the far side of the Sun with helioseismic holography.

Active regions on the near side can clearly be seen in Lyman $\alpha$ observations of the Sun. This is due to the fact that active regions are much brighter in Lyman $\alpha$ radiations than the quiet Sun. But the Lyman $\alpha$ from the Sun can not only be seen by looking at the Sun; it can also be seen by looking at secondary sources in the interplanetary medium, since the interplanetary medium contains H atoms that re-emit the Lyman $\alpha$ radiation from the Sun. As the interplanetary medium not only sees the near side of the Sun, but also the far side Lyman $\alpha$ radiation can be used to monitor active regions on the far side of the Sun. This concept was first proven by \citet{2000GeoRL..27.1331B} who used Lyman $\alpha$ observations from the SWAN instrument on SOHO to see active regions on the far side of the Sun.

There are two disadvantages of both methods. The first one is that the resolution of the active regions on the far side of the Sun is not really good -- something we cannot do anything about. The other one is that it is not really active regions that we are interested in, in order to predict the space weather. It is ratter the solar flares, as flares (and corona mass ejections) impact the Earth's environment -- this could be improved by using flare driven global oscillations. The idea is that if flares drive the high-frequency waves they will be driven equally well on the far side as on the near side of the Sun and as the waves that we have studied are low-degree oscillations we would see the signal on the near side in both cases as these waves travel almost directly though the Sun. The ability to differentiate between far side active regions and flares is important as it is not all active regions that host large flares and can thus be damaging for the Earth's environment.

The concept of using the low-degree waves to predict flares on the far side of the Sun still awaits to be proven and of course relies on the assumption that the correlation that we have seen between the solar X-ray flux and the energy in the high-frequency part of the acoustic spectrum is indeed caused by flare driven global waves, but it holds the potential to improve the precision in space weather prediction.

\section{Flare Generated Waves in other Solar-like Stars}
As the correlation between the solar X-ray flux and the energy in the high-frequency part of the acoustic spectrum was found using disk-integrated data it might be possible to see the same kind of correlation in other solar-like stars. \citet{2007MNRAS.381.1001K} has already shown that other solar-like stars very likely display waves with frequencies above the atmospheric acoustic cut-off frequency and it will thus at least be possible to analyze if correlations like those for the Sun also exist for other stars. We have great expectations for the data the we are going to receive from the Kepler mission (Metcalfe, p. ??? in these proceedings) as these will have a temporal corvage of more than 3.5 years and hopefully have a precision that allows observations of high-frequency waves in other solar-like stars. If this is the case and again assuming that the correlations that we see between the solar X-ray flux and the energy in the high-frequency part of the acoustic spectrum is caused by flare driven global waves then we should be able to observe the signature of flares on other solar-like stars. 

Such observations could be used in the quest to understand the solar dynamo as the Kepler data will hopefully allow us to observe and study stellar cycles in solar-like stars by observing the shifts in the frequencies and amplitudes that are caused by the stellar cycles \citep[see e.g.][]{2000MNRAS.313...32C}. By combining the observations of stellar cycles and stellar flares we would be able to study how this relates to models of stellar cycles. This is an important issue as it is not fully understood why the Sun, in both cycle 21 and 23, showed a two years lag between the maximum for the appearance of flares to that of the sun spot number \citep{2004SoPh..219..125V}

\acknowledgements 
I thank Hans Kjeldsen, Bill Chaplin, Doughlas Gough, Mike Thompson, J$\o$rgen Christensen-Dalsgaard, Bernhard Fleck and Hugh Hudson for useful discussion.
I thank the local organising committee for helping me with financial support to attend the meeting. I also acknowledge financial support from the Danish Natural Science Research Council and the Danish AsteroSeismology Centre.

\newpage


\begin{thebibliography}{}
\bibitem[Ambastha \& Antia(2006)]{2006SoPh..238..219A} Ambastha, A., \& 
Antia, H.~M.\ 2006, \solphys, 238, 219

\bibitem[Balmforth \& Gough(1990)]{1990ApJ...362..256B} Balmforth, N.~J., 
\& Gough, D.~O.\ 1990, \apj, 362, 256

\bibitem[Bertaux et al.(2000)]{2000GeoRL..27.1331B} Bertaux, J.-L., 
Quemerais, E., Lallement, R., Lamassoure, E., Schmidt, W., 
\& Kyr{\"o}l{\"a}, E.\ 2000, Geophys. Res. Letters, 27, 1331 

\bibitem[Brookes et al.(1976)]{1976Natur.259...92B} Brookes, J.~R., Isaak, 
G.~R., \& van der Raay, H.~B.\ 1976, \nat, 259, 92 

\bibitem[Brown et al.(1992)]{1992ApJ...394L..65B} Brown, T.~M., Bogdan, 
T.~J., Lites, B.~W., \& Thomas, J.~H.\ 1992, \apjl, 394, L65 

 \bibitem[Chaplin et al.(1995)]{1995soho....2..335C} Chaplin, W.~J., Elsworth, Y., Howe, R., Isaak, G.~R., McLeod, C.~P., Miller, B.~A., \& New, R.\ 1995, 4.~SOHO Workshop Helioseismology, Vol.~2, p.~335 - 339, 2, 335 
 
 \bibitem[Chaplin et al.(2000)]{2000MNRAS.313...32C} Chaplin, W.~J., 
Elsworth, Y., Isaak, G.~R., Miller, B.~A., 
\& New, R.\ 2000, \mnras, 313, 32 
 
 \bibitem[Chaplin et al.(2003)]{2003ESASP.517..247C} Chaplin, W.~J., 
Elsworth, Y., Isaak, G.~R., Marchenkov, K.~I., Miller, B.~A., \& New, R.\ 
2003, GONG+ 2002.~Local and Global Helioseismology: the Present and Future, 
517, 247 

\bibitem[Clark(2006)]{2006Natur.441..402C} Clark, S.\ 2006, \nat, 441, 402 
 
\bibitem[Claverie et al.(1981)]{1981Natur.293..443C} Claverie, A., Isaak, 
G.~R., McLeod, C.~P., van der Raay, H.~B., 
\& Roca Cortes, T.\ 1981, \nat, 293, 443 

\bibitem[Deubner(1995)]{1995ASPC...76..303D} Deubner, F.-L.\ 1995, GONG 
1994.~Helio- and Astro-Seismology from the Earth and Space, 76, 303 

\bibitem[Finsterle et al.(2004)]{2004SoPh..220..317F} Finsterle, W., 
Jefferies, S.~M., Cacciani, A., Rapex, P., Giebink, C., Knox, A., 
\& Dimartino, V.\ 2004, \solphys, 220, 317 

\bibitem[Foglizzo(1998)]{1998A&A...339..261F} Foglizzo, T.\ 1998, \aap, 339, 261

\bibitem[Foglizzo et 
al.(1998)]{1998A&A...330..341F} Foglizzo, T., et al.\ 1998, \aap, 330, 341

\bibitem[Garcia et al.(1998)]{1998ApJ...504L..51G} Garcia, R.~A., et al.\ 
1998, \apjl, 504, L51 

\bibitem[Gavryusev \& Gavryuseva(1999)]{1999MNRAS.303L..63G} Gavryusev, 
V.~G., \& Gavryuseva, E.~A.\ 1999, \mnras, 303, L63 

\bibitem[Goldreich 
\& Kumar(1988)]{1988ApJ...326..462G} Goldreich, P., \& Kumar, P.\ 1988, \apj, 326, 462 

\bibitem[Harvey et al.(1993)]{1993ASPC...42..111H} Harvey, J.~W., Duvall, 
T.~L., Jr., Jefferies, S.~M., 
\& Pomerantz, M.~A.\ 1993, GONG 1992.~Seismic Investigation of the Sun and Stars, 42, 111

\bibitem[Hudson(1972)]{1972SoPh...24..414H} Hudson, H.~S.\ 1972, \solphys, 
24, 414

\bibitem[Jefferies et al.(1997)]{1997ApJ...485L..49J} Jefferies, S.~M., 
Osaki, Y., Shibahashi, H., Harvey, J.~W., D'Silva, S., 
\& Duvall, T.~L., Jr.\ 1997, \apjl, 485, L49 

\bibitem[Jim{\'e}nez et al.(2005)]{2005ApJ...623.1215J} Jim{\'e}nez, A., 
Jim{\'e}nez-Reyes, S.~J., \& Garc{\'{\i}}a, R.~A.\ 2005, \apj, 623, 1215

\bibitem[Jim{\'e}nez(2006)]{2006ApJ...646.1398J} Jim{\'e}nez, A.\ 2006, 
\apj, 646, 1398

\bibitem[Karoff(2007)]{2007MNRAS.381.1001K} Karoff, C.\ 2007, \mnras, 381, 
1001 
\bibitem[Karoff \& Kjeldsen(2008)]{karoff2008} Karoff, C., \& Kjeldsen, H.\ 2008, \apjl, 678, L73 

 \bibitem[Kumar \& Lu(1991)]{1991ApJ...375L..35K} Kumar, P., \& Lu, E.\ 
1991, \apjl, 375, L35  

\bibitem[Libbrecht(1988)]{1988ApJ...334..510L} Libbrecht, K.~G.\ 1988, 
\apj, 334, 510 

\bibitem[Lindsey \& Braun(2000)]{2000Sci...287.1799L} Lindsey, C., \& Braun, D.~C.\ 2000, Science, 287, 1799 

\bibitem[Stein \& Nordlund(2001)]{2001ApJ...546..585S} Stein, R.~F., \& 
Nordlund, {\AA}.\ 2001, \apj, 546, 585 

\bibitem[Veronig et al.(2004)]{2004SoPh..219..125V} Veronig, A.~M., Temmer, 
M., \& Hanslmeier, A.\ 2004, \solphys, 219, 125 

\bibitem[Wolff(1972)]{1972ApJ...176..833W} Wolff, C.~L.\ 1972, \apj, 176, 833 

\end{thebibliography}
\end{document}